\documentclass[9pt,twocolumn,twoside]{osajnl}
\usepackage{caption}
\usepackage{subcaption}
\usepackage{multirow}
\usepackage{hyperref}

\journal{preprint} 

\setboolean{shortarticle}{true}

\title{Highly efficient DUV generation at 100~kHz via Yb pumped four-wave mixing in stretched hollow-core fibers}

\author[1,2]{Ruaridh Forbes}
\author[3]{Paul Hockett}
\author[3,*]{Rune Lausten}

\affil[1]{Department of Chemistry and Department of Physics, University of California, Davis, CA 95616, USA}
\affil[2]{Linac Coherent Light Source, SLAC National Accelerator Laboratory, Menlo Park, CA 94025, USA}
\affil[3]{National Research Council of Canada, 100 Sussex Drive, Ottawa, ON K1A 0R6, Canada}

\affil[*]{Corresponding author: rune.lausten@nrc-cnrc.gc.ca}

\begin{abstract}
We report the generation of the fifth harmonic of Yb at 206~nm with pulse energies exceeding 16~$\mu$J and durations of approximately 100~fs at a repetition rate of 100~kHz. The deep ultraviolet pulses are produced using four-wave difference frequency mixing in a He-filled stretched hollow-core fiber, driven by a pump at 343~nm and seeded at 1030~nm. Guided by simulations, we carefully optimize the process, resulting in a conversion efficiency of $\sim$30\% from the 343~nm pump beam.
\end{abstract}

\setboolean{displaycopyright}{true}

\begin{document}


\maketitle

The efficient generation of femtosecond pulses in the vacuum ultraviolet (VUV, 100–200 nm) and deep ultraviolet (DUV, 200–280 nm) spectral regions remains a longstanding challenge in nonlinear optics~\cite{Chapman2014f}. Despite their utility across a wide range of applications — including ultrafast optics, molecular spectroscopy, and dynamics~\cite{Eikema2011, Chergui2014b, Reid2016b, schuurman2022TimeresolvedPhotoelectronSpectroscopy}, as well as VUV comb spectroscopy ~\cite{Picqu__2019_Freq_Comb_spectroscopy, Cing_z_2012_Cav_HHG_freq_Comb_Th} — these pulses are difficult to produce because of the limited transparency and phase-matching capabilities of nonlinear crystals in this wavelength range. 

Although nonlinear crystals have been successfully employed in some cases, their performance in the DUV/VUV is often constrained by absorption, photoinduced damage, and the eventual need for mechanical translation or replacement due to color center formation. As an alternative, gas-based nonlinear optics has emerged as a promising route for UV frequency conversion, leveraging noble gases’ wide transparency range and smooth dispersion properties.  Recent demonstrations include DUV generation at 267~nm with $\sim$13\% efficiency and at shorter wavelengths (224–240~nm) with up to $\sim$3.8\% using argon-filled hollow-core fibers (HCFs) \cite{Jailaubekov_2005}. Even higher FWM efficiencies, up to 38\% at 270~nm, have been reported using kagomé-style photonic crystal fibers (kagomé-PCF) \cite{belli_2019}, while femtosecond VUV pulses at 160~nm (7.75~eV) have been produced with $\sim$10\% efficiency in helium-filled stretched HCFs \cite{Forbes_2024_FWM_HCF_VUV_5w}. (Note that herein that we use HCF throughout to refer to capillary type fibers, and PCF to refer to microstructured fibers.)

The choice of fiber type significantly influences scalability: kagomé and anti-resonant fibers offer lower pump energy thresholds, enabling high-repetition-rate operation, but are limited by annulus resonances and complex dispersion profiles \cite{Zeisberger_2018_ARF_dispersion}. In contrast, stretched HCFs offer excellent modal control and phase-matching capabilities, enabling high-efficiency frequency conversion when paired with noble gases. This is particularly valuable for advanced applications that require bright, tunable, and ultrashort pulses in the DUV and VUV, driven with higher pump energies and lower repetition rates, e.g. $\mu$J drivers in the kHz to hundred-kHz range.

\begin{figure}[t]
  \centering
    \includegraphics[width=0.4\textwidth]{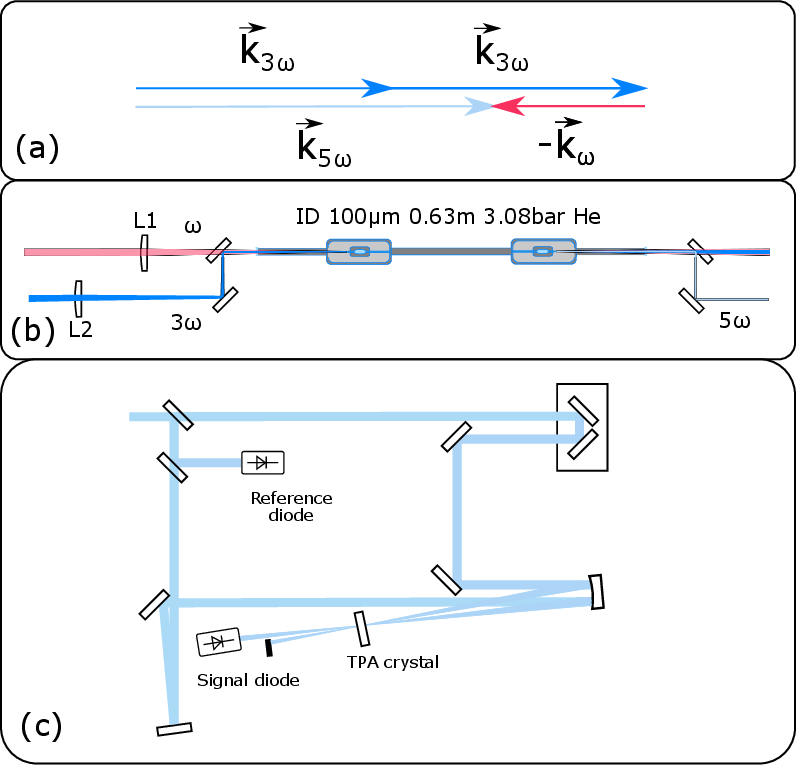}
  \caption{\label{fig:5w_generation}(a) Wave vector diagram for the phase matching between the $\omega$ and 3$\omega$ beams in the FWDFM process. (b) Experimental setup of the HCF used to generate the fifth harmonic. The fundamental, and third harmonic beams are recombined on a dichroic mirror (DM), mode-matched to the fiber EH$_{11}$ mode. L1 is a thin fused silica lens with f=250~mm, and L2 is a thin CaF$_{2}$ lens with f=450~mm. The fiber chamber is filled with He to the phase matching pressure of 3.08~bar. The fifth harmonic is separated from the driving beams with dielectric mirrors, and either sent to a spectrometer (MayaPro, Ocean-Optics), or to a powerhead (3A-FS, Ophir Optics). (c) The setup for measuring the pulse length of the DUV pulse, though two-photon absorption autocorrelation in CaF$_{2}$. Details are in the main text.}
\end{figure}

In the following we build on foundational work on nonlinear optics in HCFs~\cite{Durfee_1997,Durfee_1999,Durfee_2002,Misoguti_2001,Tzankov_2007} and report the generation of the fifth harmonic (206~nm, 6.02~eV) using a high-repetition rate (100~kHz) Yb-based femtosecond laser to drive four-wave difference frequency mixing (FWDFM)  (see Fig.~\ref{fig:5w_generation}(b)) in helium-filled stretched HCF. Using a 343~nm pump and 1030~nm seed, we achieve $>$16~$\mu$J pulse energies and $\sim$30\% conversion efficiency. Fig.~\ref{fig:5w_generation}(a) illustrates the mixing scheme.

In the presented work, two key components contributed to the significant improvement in frequency conversion efficiency. The first was the use of stretched HCFs, following the fabrication guidelines presented by Nagy \textit{et al.}~\cite{Nagy_2008}. The second key component was the use of the modeling software Luna.jl~\cite{Luna}, which has been shown to be quantitatively predictive in simulating nonlinear interactions in HCFs across various measurements~\cite{Brahms_2019_HE_UV_RDW_compact,Grigorova_2023_DispTuning_NL_pulseDynamics, Forbes_2024_FWM_HCF_VUV_5w, Kumar_2025_Luna_use_CH4_filled_HCF,Brahms2023Efficient}. Together, these components represent a significant step toward compact, high-flux DUV/VUV sources for future ultrafast photonics applications.

The pump laser is a Yb based femtosecond amplifier (Carbide CB3-80W, from Light Conversion) with a wavelength of 1030~nm, and pulse length of 170~fs. The fundamental pulse energy of this system is 800~$\mu$J, and the repetition rate is 100~kHz. The pulse is split into a part for the third harmonic generation setup, and a seed pulse, with the remainder ($\sim$ 40~W) sent to a water-cooled power head outside the experimental setup. The power of the seed and third harmonic pump can be varied independently by half waveplates and TFPs in both arms: in the reported experiments, the pulse energy of the seed and the pump were varied between 0 and 70~$\mu$J.
Fig.~\ref{fig:5w_generation}(b) illustrates the HCF setup. The seed and the pump are coupled to a 0.63~m long stretched HCF~\cite{Nagy_2008}, with an ID=100~$\mu$m (Polymicro, TSP100665), through a 1.0~mm MgF$_2$ entrance window (Crystran, VUV grade). The fiber can be evacuated and filled with the noble gas He. The experimental coupling efficiencies achieved were 90$\%$ and 89$\%$ for the pump and the seed, respectively. The coupling efficiency is defined as the fraction of the light coupled into the fiber (input divided by output, and corrected by the theoretical HCF transmission).
%

After the fiber, the DUV pulse is separated from the pump and seed with dielectric mirrors (45~deg, 206~nm HR, Light Conversion), and either sent to a spectrometer (MayaPro, Ocean Optics), or to a thermal power head (3A-FS Ophir), which allowed us to measure the DUV power, or a pulse autocorrelator (see Fig.~\ref{fig:5w_generation}(c)). In the quoted results from the power measurements, we correct for the reflectivity of the dielectric, and the reflection losses of the exit window in the beam path, using reflectivity curves from the manufacturer and refractive indices from literature~\cite{RefractiveIndex_info}. The DUV pulse generated from the capillary at optimal phase matching pressure showed good stability. Short term we see DUV pulse energy fluctuations of $\leq$0.5$\%$, and the average DUV pulse energy for longer intervals of several hours remains constant, provided that beam pointing into the capillary remains stable and the vacuum system prevents air leakage into the beam path. After cleaning the HCF chamber and all internal parts with standard Ultra High Vacuum (UHV) techniques, there are no signs of photochemistry leading to deposition on the windows or inside the HCF, and issues with color center formation in the input and exit windows do not appear to be an issue with VUV grade MgF$_2$, at $\sim$3~mm spot size for 5$\omega$ on the exit window.
This statement holds for the duration of the current experiments, conducted over several weeks and with 10's hours of run-time at various input fluences, of which $<$10~hours run-time were at the highest pump energies; ongoing work is probing longer-term stability at high energies.
Considering that the wave vectors mismatch, $\Delta k$, of the FWDFM process, can be written as $\Delta k = 2k_{3\omega}-k_\omega-k_{5\omega}$, and inserting the expressions for the contributions to the wave vectors from the dispersion of the noble gas, at pressure $P$, and the modal dispersion from the capillary propagation, based on the Marcatili and Schmeltzer paper \cite{Marcatili}, the wave vector mismatch splits into a positive, pressure dependent, gas term, and a negative modal propagation term $\Delta k = P \Delta k_{gas}-\Delta k_{mode}$ \cite{Durfee_2002}. 
To characterize the pressure dependence of the 5$\omega$ conversion efficiency, pulse energy measurements of the DUV were performed as a function of He pressure, for set pump and seed powers. Additionally, the power scaling of the DUV generation was investigated by varying the seed and pump power, while maintaining the fiber at the optimal phase matching pressure.
The experimental measurements may be compared with the simple theory expression: 
\begin{equation}
I_{5\omega} \propto  N^2 \mid {\chi}^{(3)} \mid^{2} \frac{I_{\omega} I_{3\omega}^2 L^2 ~\mbox{sinc}^2(\Delta k L/2)}{n_{\omega} n_{3\omega}^2 \lambda_{5\omega}^2 } \\
\label{eq:simple-PM}
\end{equation}
Previous work from Ghotbi \textit{et al.} utilized this expression to obtain excellent agreement with FWM experiments to generate the fifth harmonic (5$\omega$) of Ti-Sapphire \cite{Noack_2muJ}.
It contains the phase-mismatch, ${\Delta}k$, $N$ (which is proportional to the pressure, $P$) is the number density  of the He atoms, ${\chi}^3$ is the third-order nonlinear coefficient, $I$ is the intensity, $L$ is the interaction length, and $n$ is the refractive index.  
The experimental power-scaling measurements agree very well with the expectations based on Eq.~\ref{eq:simple-PM}, demonstrating linear scaling with the seed and quadratic with the pump, over the range explored, which was limited by non-linear effects, such as self-focusing and color center formation in the input window. In the current experiments, pump and seed powers up to 58~$\mu$J resulted in DUV pulses of $\sim$16~$\mu$J (corresponding to a conversion efficiency of $\sim$30$\%$).


\begin{figure}[hbt]
         \centering
         \includegraphics[width=0.45\textwidth]{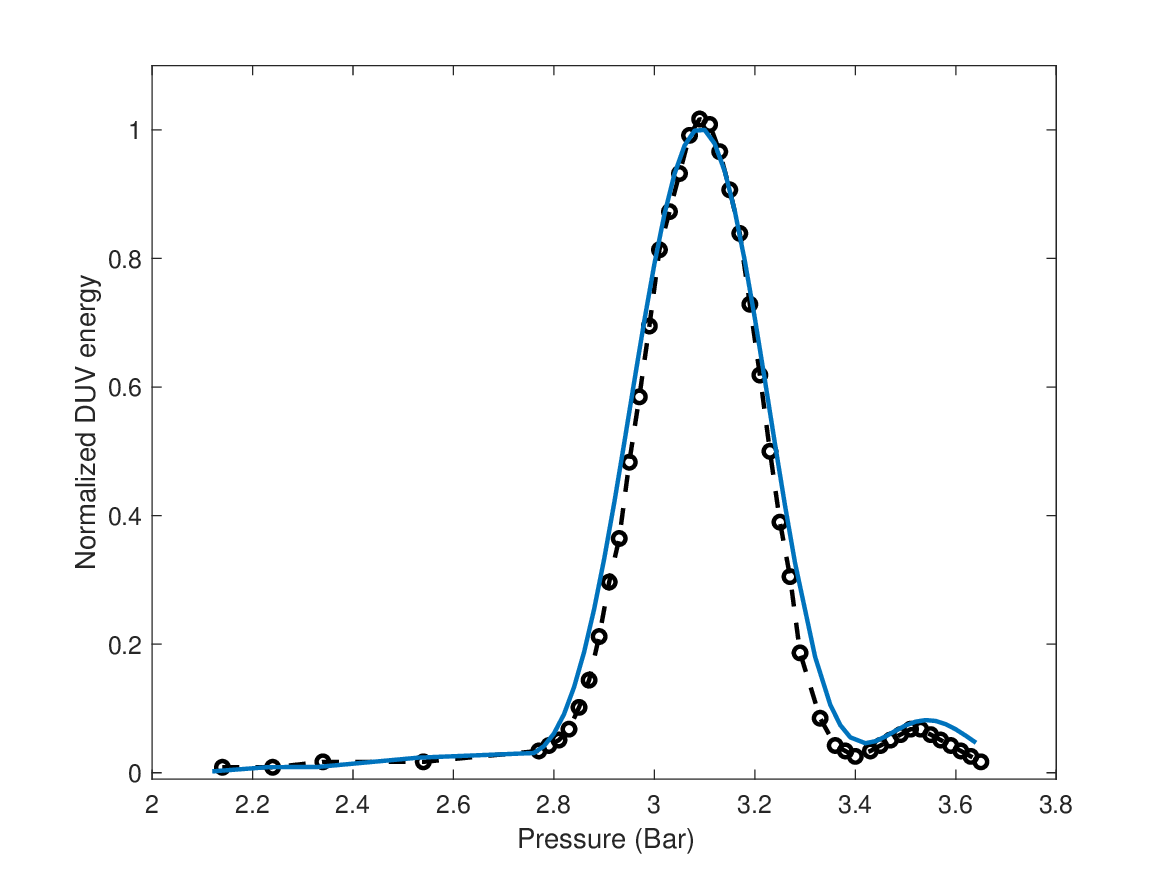}
         \caption{Experimental phase-matching curve (circles with dashed line), with the simulation results from Eq.~\ref{eq:simple-PM} overlay (solid line). Both the calculated curve and the experimental data are normalized.}
         \label{fig:5w_phasematching_curve}
\end{figure}

As shown in Fig.~\ref{fig:5w_phasematching_curve}, the experimental pressure-dependence measurements also agree very well with the expectations of the simple theory (see Eq.~\ref{eq:simple-PM}) An optimized phase matching pressure of 3.08~bar of He is found for the employed parameters. The theory phase-matching curve was based on Eq.~\ref{eq:simple-PM} above, using the Sellmeier formula for He from Ermolov {\it et al.} \cite{Ermolov2015}. This theory expression reproduces both the position and width of the main peak of the phase-matching curve quite well, along with the smaller peaks at higher and lower pressures, which are expected from the $sinc^2(\Delta k L/2)$ phase-matching function. 

In order to be able to model the process in even more detail, and also include effects of cross-phase modulation (XPM), self-phase modulation (SPM), ionization, modal coupling, and pump depletion, we used Luna.jl~\cite{Luna}, a flexible platform for the simulation of nonlinear optical dynamics in waveguides, which has been shown to be quantitatively predictive against experiment \cite{Travers2019}. Using Luna.jl to generate a phase matching curve for the experimental conditions produces a curve that is virtually identical to the simple theory expression.
This observation is consistent with the expectation that, at these pump/seed powers, the process is well described by the simple model: there is no significant SPM, XPM, or plasma interaction at the fiber entrance that would otherwise initiate mode coupling or modify the driving fields.

\begin{figure}[tbh]
         \centering
         \includegraphics[width=0.45\textwidth]{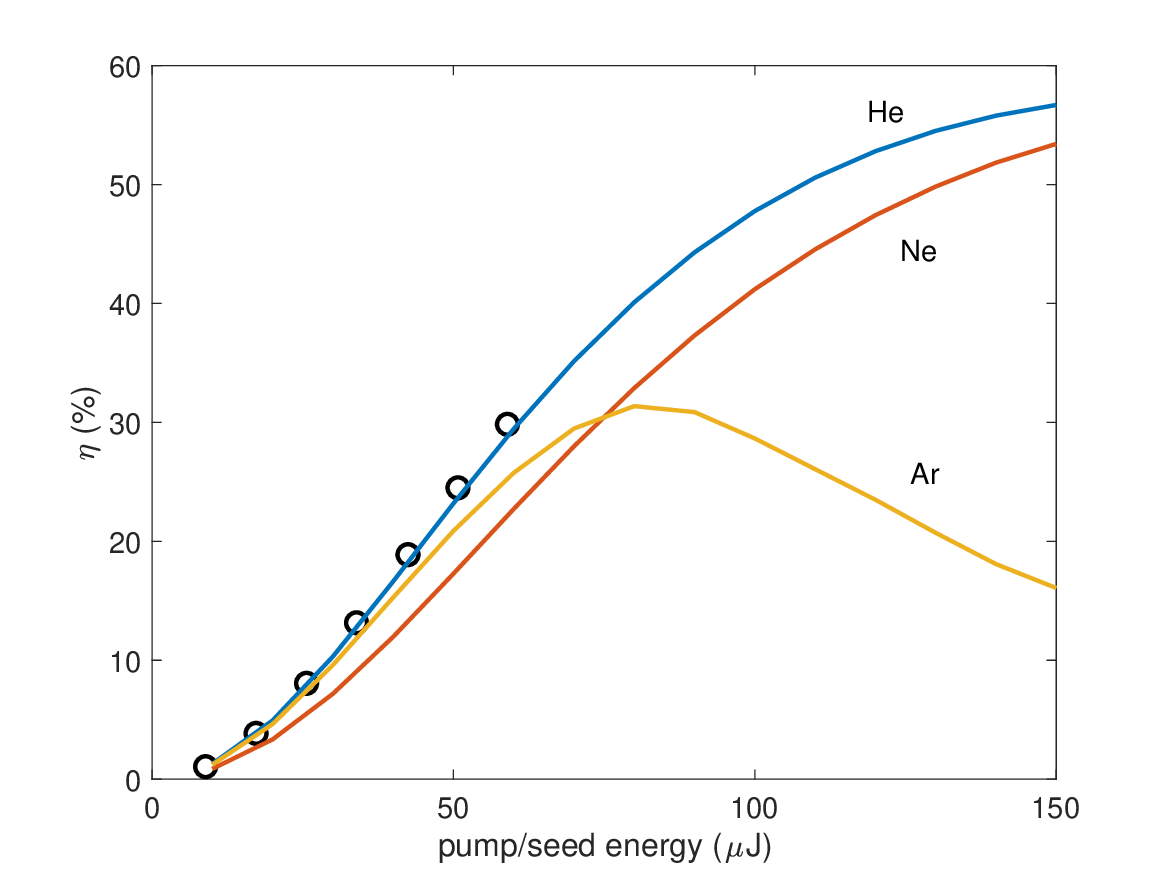}
         \caption{Conversion efficiency, $\eta$, of the 343~nm pump into DUV as a function of the pump/seed power for the fixed HCF geometry used. Circles are the experimental points, recorded with an He filled HCF. Pump and seed energies were varied together.}
         \label{fig:Luna_5w_scaling_Yb}
\end{figure}


Fig.~\ref{fig:Luna_5w_scaling_Yb} shows experimental power scaling and Luna.jl results. The results are in good agreement for the current experiments with an optimally phase-matched He filled HCF. Exploring the power scaling beyond the current experiment, the Luna.jl simulations show no significant detrimental effects even when pumping/seeding with 125~$\mu$J (pump and seed energies are varied together in these simulations). At these powers, the simulations predict the generation of $\sim$45~$\mu$J of 206~nm, corresponding to a conversion efficiency of $\sim$50$\%$ when the process is optimally driven; see Fig.~\ref{fig:Luna_5w_scaling_Yb}. From this point on, the conversion efficiency starts to roll over mainly due to pump depletion.
The pump depletion is accompanied by temporal/spectral broadening of the DUV pulse, eventually resulting in a spectrum with a dip in the middle. 

 It is interesting to compare the different noble gases as the non-linear medium. Performing the same simulation for neon and argon, at their optimal phase matching pressure (1.39~bar and 0.127~bar, respectively) produces the results shown by the two other curves in Fig.~\ref{fig:Luna_5w_scaling_Yb}.

The main observations when comparing He with Ne and Ar is that the intensities where the FWDFM process is optimally driven moves down, as does the maximal conversion efficiency ($\sim$50$\%$ for He, $\sim$45$\%$ for Ne and $\sim$28$\%$ for Ar). 
By running the simulation with the ionization term switched off, it is clear that there is no sign of plasma interaction for He and Ne, but that the roll over for Ar can be attributed to this.
Given the Ionization Potential (IP) of He=24.59~eV, Ne=21.56~eV and Ar=15.76~eV ~\cite{NIST_ASD}, it is clear that the lowest three-photon process that could lead to ionization is 5$\omega$+5$\omega$+5$\omega$=15$\omega$ (18.06~eV). Three-photon ionization with 5$\omega$ is thus able to ionize Ar, but not He or Ne. This qualitatively explains the shape of the curves we see in Fig.~\ref{fig:Luna_5w_scaling_Yb}. As the intensity 5$\omega$ increases, it eventually leads to ionization for the case of Ar, which in turn spoils the conversion. 




These simulations provide interesting insights into the more complicated optimally and strongly-driven regimes.
The presentation of full simulation results is beyond the scope of this letter.
\begin{figure}[t]
  \centering\begin{subfigure}[]{0.4\textwidth}
     \centering
     \includegraphics[width=\textwidth]{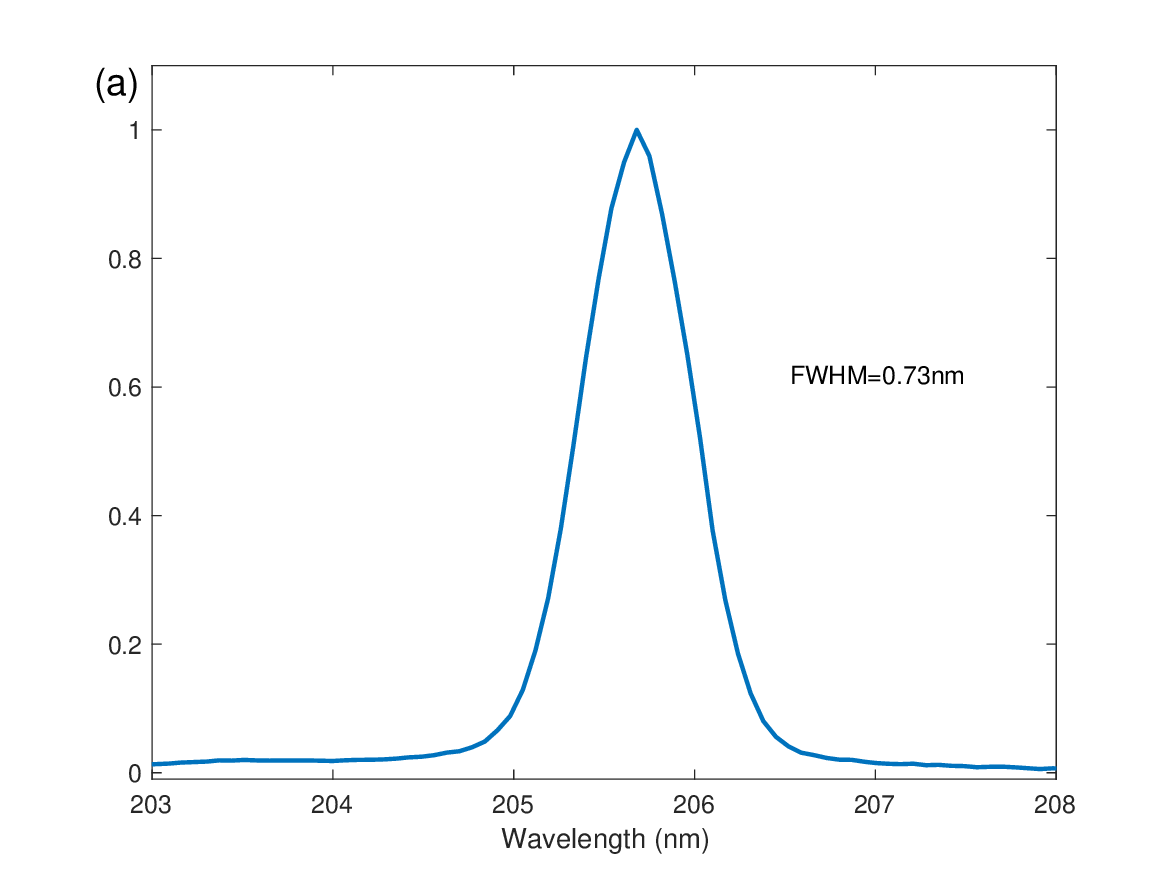}
    \end{subfigure}
    
    \begin{subfigure}[]{0.4\textwidth}
     \centering
     \includegraphics[width=\textwidth]{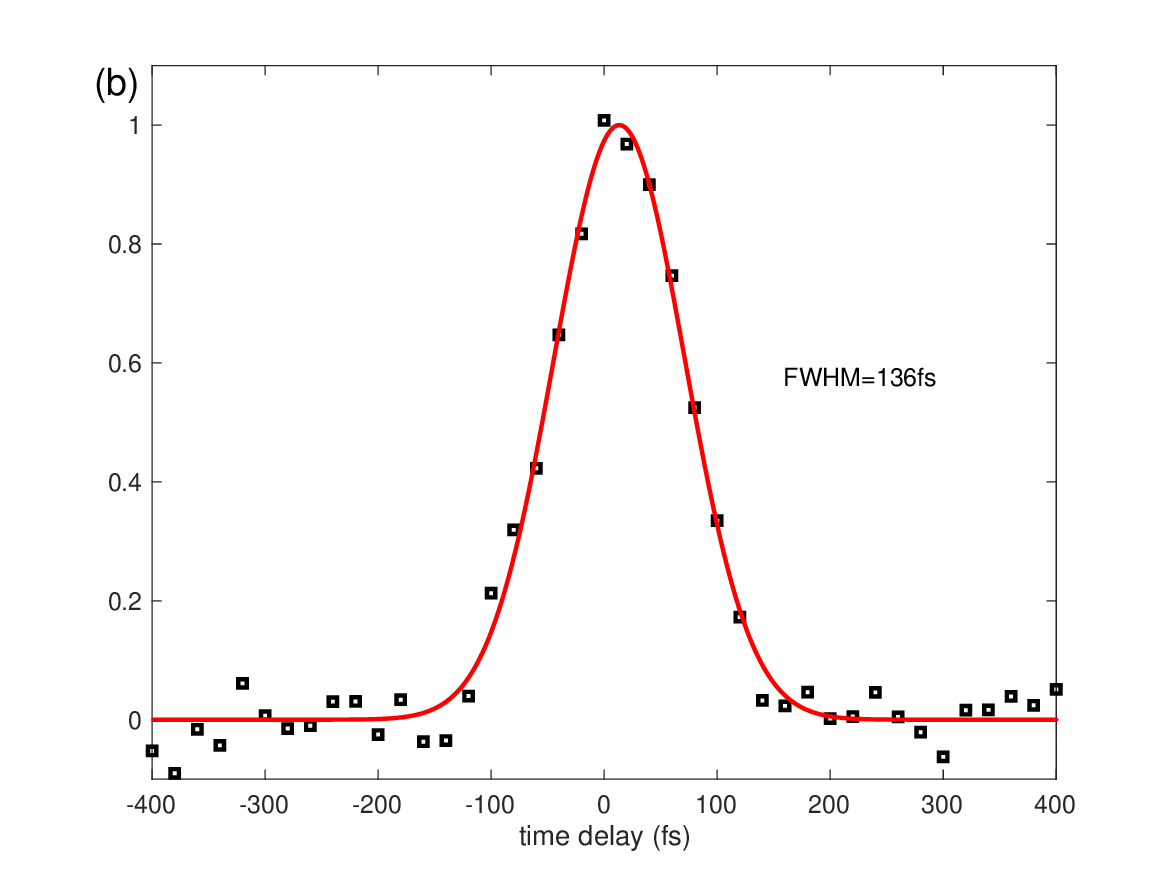}
   \end{subfigure}
    \caption{\label{fig:5w_spectrum_CC}(a) Measured spectrum (Ocean Optics MayaPro) of the generated DUV pulses at 206~nm, with an full width at half maximum (FWHM) of 0.73~nm, corresponding to a transform-limited pulse length of 85~fs. (b) Two-photon absorption autocorrelation trace of the DUV pulse, measured in CaF$_{2}$. The FWHM of 136~fs corresponds to a deconvolved pulse duration of 96~fs.}
\end{figure} 
The temporal and spectral characterization of the DUV pulses was performed by separate but concurrent measurements. A typical spectrum of the DUV pulse at 3.08~bar of He is shown in Fig.~\ref{fig:5w_spectrum_CC}(a). The full width at half maximum (FWHM) of the spectrum is $\sim$ 0.73~nm, which corresponds to a transform limited pulse length of 85~fs. The temporal measurement was performed by two-photon absorption autocorrelation in CaF$_2$~\cite{Homann_2011},  where the DUV beam split into a strong pump and a weak probe, using a CaF$_2$ window, and then both beams are focused in a 2~mm CaF$_2$ window with a concave Al mirror in a non-collinear geometry. Fig.~\ref{fig:5w_generation}(c) illustrates this measurement scheme.
Monitoring both the transmitted and reflected seed on fast photodiodes, we can measure a dip in transmission of the seed as a function of the pump delay. Typical cross correlation traces at 10-100~nJ pulse energy result in autocorrelation traces with a FWHM width of ~136~fs, corresponding to a pulse length of 96~fs (see Fig.~\ref{fig:5w_spectrum_CC}(b).

 This result is in excellent agreement with expectations, since propagation of a transform limited 90~fs pulse at 206~nm through the 2~mm MgF$_2$ exit window of the HCF chamber, and 3~mm of CaF$_2$ in the TPA autocorrelatior (2mm CaF$_2$ window for beam splitter, and 1~mm for the propagation through half of the CaF$_2$ window used for TPA)  would stretch it to $\sim$~95~fs.

In conclusion, we have demonstrated the generation of near transform limited 96~fs VUV pulses at 206~nm, with pulse energies as high as 16~$\mu$J, which corresponds to a conversion efficiency of $\sim$30$\%$. The source is pumped by a commercial femtosecond Yb amplifier and uses a relatively simple experimental setup, taking advantage of phase-matched FWDFW in a He-filled stretched HCF. Separation of the DUV beam from the driving beams is achieved using dielectric mirrors. The high conversion efficiency of this source, makes it an ideal path to femtosecond tuneable DUV pulse generation, where seeding with signal or idler range of a standard OPA would provide tunability from 234 to 208~nm (5.3-5.96~eV)~\cite{Noack_tuneableVUV_45fs,Trabs_2014,Forbes2021}. 
An extension of this work into VUV is underway, to demonstrate generation of 6$\omega$ (172nm, 7.21~eV) and 7$\omega$ (147~nm, 8.43~eV) in a single HCF pumped by an Yb amplifier.
The combination of brightness, short pulse duration, stability, and high photon energy makes this source attractive for many forms of photon-in, electron-out spectroscopies~\cite{Trabs_2014,Forbes2018,Forbes2021}.  
The DUV pulse could be pre-compensated for propagation through the experimental chamber input window, by adding positive dispersion to the seed~\cite{Noack_12fs,Suzuki_Spectra_Phase_Transfer_UV, Zhang_2024_FWM_PhaseTransfer}, and it should be possible to make even shorter DUV pulses by seeding with a broader spectrum \cite{Noack_12fs}. Finally, this technique also lends itself to the direct generation of DUV pulses in desired polarization states, e.g. circular, by controlling the polarization of the driving fields~\cite{Lekosiotis_2020}.






\begin{backmatter}
\bmsection{Funding} R.F. acknowledges support from the Linac Coherent Light Source (SLAC National Accelerator Laboratory), supported by the U.S. Department of Energy, Office of Science, Office of Basic Energy Sciences (Contract No. DE-AC02-76SF00515). P.H. and R.L. acknowledge support from the NRC-CSTIP Quantum Sensors program (QSP-075-1).

\bmsection{Acknowledgments} We thank Light Conversion for providing the CARBIDE laser and acknowledge insightful discussions with Ignas Stasevičius and Lukas Rimkus. We also thank Christian Brahms and John Travers (Heriot-Watt University) for discussions on the numerical modeling using Luna.jl.  

\bmsection{Disclosures} The authors declare no conflicts of interest.

\bmsection{Data availability} Data underlying the results presented in this paper are available from the authors with reasonable request.
\end{backmatter}

\bigskip
\bibliography{sample, roadmaps_and_reviews_071223}
%
%

\end{document}